\documentclass[aps,prl,amsmath,amssymb,amsfonts,showpacs,superscriptaddress,floatfix,twocolumn]{revtex4}

\usepackage{graphicx}
\usepackage{epsfig}
\usepackage[english]{babel}
\def\one{\ensuremath{\hbox{$\mathrm I$\kern-.6em$\mathrm 1$}}}

\begin{document}

\title{Continuous Matrix Product States for Quantum Fields}

 \author{F. \surname{Verstraete}}
 \affiliation{University of Vienna, Faculty of Physics, Boltzmanngasse 5, 1090 Wien, Austria}
 \author{J.~I. \surname{Cirac}}
 \affiliation{Max-Planck-Institut f\"ur Quantenoptik, Hans-Kopfermann-Str. 1, Garching, D-85748, Germany}

\begin{abstract}
We define matrix product states in the continuum limit, without any reference to an underlying lattice parameter. This allows to extend the density matrix renormalization group and variational matrix product state formalism to quantum field theories and continuum models in 1 spatial dimension. We illustrate our procedure with the Lieb-Liniger model.

\end{abstract}

\maketitle

The numerical renormalization group (NRG) of Wilson \cite{Wilson} and the density matrix renormalization group (DMRG) of White \cite{White} revolutionized the way strongly correlated quantum systems can be simulated and understood. The applicability of those approaches has been better understood during the last 5 years by rephrasing those methods in terms of matrix products states (MPS) \cite{MPS,reviewMPS}; the success of NRG and DMRG relies on the fact that those MPS give a very accurate description of the correlations and entanglement present in ground states of 1-D quantum spin systems \cite{VC06,HastingsAREA}. This insight led to several important extensions of DMRG, as MPS can also be used to describe dynamical properties \cite{dynamicsMPS} and can be used as a stepping stone for constructing higher-dimensional analogues known as projected entangled pair states (PEPS) \cite{PEPS}.

In this paper, we show how this formalism of MPS can be adopted to describe quantum field theories. We will define a new family of states that we call continuous MPS (cMPS) that describe field theories in 1 spatial dimension. We will also show that cMPS can be understood as the continuous limit of standard MPS. Those cMPS can be used as variational states for finding ground states of quantum field theories, as well as to describe real-time dynamical features. Just as MPS capture the entanglement structure of low-energy states of quantum spin systems, the cMPS seem to capture the entanglement features of the low-energy states of quantum field theories. We will illustrate this on the hand of simulations that we have done on the Lieb-Liniger model \cite{LiebLiniger} which describes a system of bosons in a one dimension interacting via a delta-potential; using cMPS with a very low bond dimension, the ground state energy density is already reproduced with extremely good precision. We will also show how one can calculate other interesting physical quantities, like correlation functions or the static structure factor.

Let us next define the cMPS, which is most easily done in the formalism of second quantization. We will consider a one-dimensional system of bosons or fermions on a ring of length $L$ and associated field operators $\hat{\psi}(x)$ with canonical commutation relations, $[\hat{\psi}(x),\hat{\psi}(y)^\dagger]_{\pm}=\delta(x-y)$ with $0\leq x,y\leq L$  space coordinates.  A cMPS is  defined as

\[|\chi\rangle={\rm Tr}_{aux}\left[\mathcal{P}e^{\int_0^L dx \left[Q(x)\otimes\openone+R(x)\otimes \hat{\psi}^\dagger(x)\right]}\right]|\Omega\rangle\]
with $Q(x),R(x)$ position dependent matrices of dimension $D\times D$ that act on a $D$-dimensional auxiliary system, $\mathcal{P}\exp$ the notation for the path-ordered exponential, ${\rm Tr}_{aux}$ the trace over the auxiliary system, and $|\Omega\rangle$ the vacuum state [$\hat{\psi}(x)|\Omega\rangle=0$]. A translational invariant state can easily be obtained by choosing $Q(x)$ and $R(x)$ independent of $x$, and a system with open boundary conditions can be obtained by replacing the ${\rm Tr}_{aux}$ by a left and right multiplication of the auxiliary system with a row and a column vector, respectively.

As we will show later, cMPS appear very naturally as a continuous limit of MPS. Thus, they automatically inherit all the properties of MPS, like the
fact that the entanglement entropy of a contiguous block of bosons is bounded above by $2\log_2(D)$. In general, the state $|\chi\rangle$ is a superposition of states with different particle number.
For the case of fermions, it is easy to enforce an occupation number with a fixed parity by introducing a $Z_2$ symmetry by choosing $Q$  and $R$ block diagonal:

\[
Q(x)=\left(\begin{array}{cc} Q_0(x) & 0\\0 & Q_1(x)\end{array}\right)\hspace{.5cm}
R(x)=\left(\begin{array}{cc} 0 & R_0(x) \\ R_1(x) & 0\end{array}\right)
\]
As a consequence, expectation values of the form $\langle \psi(x)^\dagger \psi(y)\rangle$ can be calculated without the need for introducing string-order like operators.

To get some intuition about the structure of such states, it is instructive to write down explicitly
\[
|\chi\rangle= \sum_{n=0}^\infty \int_{0<x_1<\ldots <x_n<L} dx_1\ldots dx_n \phi_n\hat{\psi}(x_1)
\ldots \hat{\psi}(x_n) |\Omega\rangle,
\]
where
\[\phi_n=
{\rm Tr}_{aux}\left[u_Q(x_1,0)R u_Q(x_2,x_1)R \ldots R u_Q(L,x_n) \right]
\]
and $u_Q(y,x)=\mathcal{P} \exp\left[\int_x^y Q(x) dx\right]$. One can interpret $u_Q$ as a free propagator, while $R$ can be understood  as a scattering matrix that creates a physical particle. In general, the MPS formalism can indeed be rephrased as a representation of  scattering events that happen in the physical vacuum of the interacting many-body state.

With the help of this definition, it is straightforward to express the norm and expectation value of operators in terms of the matrices $R$ and $Q$.
For the sake of simplicity, we will consider a bosonic system and assume translational invariance. Note that for inhomogeneous systems one can proceed in a very similar way. Using the commutation relations of the field operators one readily finds $\langle\chi|\chi\rangle = {\rm Tr} \left(e^{T L}\right)$ and
 \begin{eqnarray}
 \langle\hat{\psi}(x)^\dagger \hat{\psi}(x) \rangle &=& {\rm Tr} \left[e^{T L} (R\otimes \bar R)\right],\nonumber \\
 \langle \hat{\psi}(x)^\dagger \hat{\psi}(0)^\dagger\hat{\psi}(0) \hat{\psi}(x) \rangle &=& {\rm Tr} \left[e^{T (L-x)} (R\otimes \bar R)e^{Tx} (R\otimes \bar R) \right],\nonumber \\
 \langle\hat{\psi}(x)^\dagger\left[-\frac{d^2}{dx^2}\right] \hat{\psi}(x) \rangle &=& {\rm Tr} \left[e^{T L} ([Q,R]\otimes [\bar Q,\bar R])\right],\nonumber
 \end{eqnarray}
where $T=Q\otimes\openone + \openone\otimes \bar Q + R\otimes \bar R$ (the bar indicates complex conjugation).  The state $|\chi\rangle$ is invariant under the ''gauge'' transformation $Q\to XQX^{-1}$, $R\to XRX^{-1}$ for arbitrary invertible $X$. This allows us to fix a gauge by imposing $Q+Q^\dagger + R^\dagger R =0$, so that we can write
\[ Q=-\frac{1}{2} R^\dagger R - i H \]
where $H=H^\dagger$ and $R^\dagger R$ is diagonal. Making the transformation $X\otimes \bar Y |a,b\rangle \to X|a\rangle\langle \bar b|Y^\dagger$, $T$ is transformed into a superoperator $\tilde T$ (mapping matrices into matrices), and we obtain that $\rho(x):=e^{\tilde Tx} \rho$ satisfies a master equation in the Lindblad form
 \[ \frac{d}{dx}\rho(x) =  -i [\tilde H,\rho(x)] + R\rho(x)R^\dagger - \frac{1}{2}\left[R^\dagger R ,\rho(x)\right]_+.\]
As a consequence,  all eigenvalues of ${\tilde T}$ have a non-positive real part, which implies that all the above quantities are well behaved in the thermodynamical limit $L\to\infty$. In a generic case, the master equation will have a unique steady state $\rho_{ss}\geq 0$, which can be chosen with unit trace. In such a case, the above expressions considerable simplify in the thermodynamic limit, since $\langle\chi|\chi\rangle={\rm Tr}(\rho_{ss})=1$,
 \begin{eqnarray}
 \langle\hat{\psi}(x)^\dagger \hat{\psi}(x) \rangle &=& {\rm Tr} \left[R^\dagger R\rho_{ss}\right],\nonumber \\
 \langle\hat{\psi}(0)^\dagger \hat{\psi}(x)^\dagger\hat{\psi}(x) \hat{\psi}(0) \rangle &=& {\rm Tr} \left[(Re^{\tilde T x}(R\rho_{ss} R^\dagger) R^\dagger \right],\nonumber \\
 \langle\hat{\psi}(x)^\dagger\left[-\frac{d^2}{dx^2}\right] \hat{\psi}(x) \rangle &=& {\rm Tr} \left[([Q,R])^\dagger [Q,R] \rho_{ss}\right],
 \end{eqnarray}
Other quantities can be similarly calculated. Note that observables defined as Fourier transforms of correlation functions, like the static structure factor, can be directly calculated in terms of the super-operator $(T-ik)^{-1}$.

In the case of a system with open boundary conditions, the eigenvalues of the matrix $\rho(x)$ would exactly correspond to the squares of the Schmidt coefficients when considering a bipartition at site $x$. This can in its turn be used to calculate the entanglement entropy of the reduced density matrices defined on given intervals.
Just as in the case of quantum spin systems, the justification for using cMPS should stem from the fact that an area law is satisfied for this entanglement entropy, eventually with logarithmic corrections in the case of critical systems.  It seems indeed possible to generalize the work of Hastings \cite{HastingsAREA} (proving the area law for 1-dimensional gapped spin systems) to the current continuous setting \cite{inpreparation}.

Let us next show how these continuous MPS can be understood as a limit of a family of MPS. For simplicity, we will consider a translational invariant system of bosons on a ring of length $L$; an identical construction works for the fermionic case. We define a family of translational invariant MPS of $N=L/\epsilon$ modes  on a discretized lattice with lattice parameter $\epsilon$ with modes $a_i$ that obey the commutation relations $\left[\hat{a}_i^\dagger,\hat{a}_j\right]=\delta_{ij}$:

\[|\chi_\epsilon\rangle=\sum_{i_1\cdots i_N}{\rm Tr}\left[A^{i_1}\cdots A^{i_N}\right] \left(\hat{\psi}_1^\dagger\right)^{i_1}\cdots \left(\hat{\psi}_N^\dagger \right)^{i_N} |\Omega\rangle\]

\begin{eqnarray*}
A^0&=& \openone +\epsilon Q\\
A^1&=&\epsilon R\\
A^n&=&\epsilon^n R^n/n!\\
\hat{\psi}_i&=&\frac{\hat{a}_i}{\sqrt{\epsilon}}
\end{eqnarray*}

Again, $|\Omega\rangle$ is the empty vacuum on which the operators $\hat{a}_i$ act ($\hat{a}_i|\Omega\rangle=0$), and we use the convention that $\left(\hat{\psi}_k^\dagger\right)^{0}=\openone$. The operators $\hat{\psi}_i$ are defined as rescaled annihilation operators ; obviously, those will become the field operators in the limit $\epsilon\rightarrow 0$:

\begin{eqnarray*} [\hat{\psi}_i,\hat{\psi}_j^\dagger]&=&\frac{\delta_{ij}}{\epsilon}\\
&\rightarrow& [\hat{\psi}(x),\hat{\psi}(y)^\dagger]=\delta(x-y).
\end{eqnarray*}
 $Q$ and $R$ are $D\times D$ matrices, and the scaling of the matrices $A^i$ as a function of $\epsilon$ has been chosen such that this limit is well defined. The matrices $A^k$ for higher $k$ have been determined by the requirement that e.g. a doubly occupied site yields the same physics as 2 bosons on 2 neighboring sites in the limit $\epsilon\rightarrow 0$. With this convention, the continuum limit of this MPS is equivalent to the continuous MPS defined before. It can also be checked that all divergencies in $1/\epsilon$ magically cancel each other, such as occurring in the case of calculating the kinetic energy

\[E_{kin}=L\langle\chi| \left(\frac{\hat{\psi}_{i+1}^\dagger-\hat{\psi}_{i}^\dagger}{\epsilon}\right)\left(\frac{\hat{\psi}_{i+1}-\hat{\psi}_{i}}{\epsilon}\right)|\chi\rangle.\]
The cancellation of the divergent terms $1/\epsilon^2$ and $1/\epsilon$ can  easily be proven by expanding
  \begin{eqnarray*}
& \sum_{\alpha,\alpha',\beta,\beta'}A^\alpha A^\beta\otimes \bar{A}^{\alpha'}\bar{A}^{\beta'}.\\
 &\langle\Omega|\psi_{i+1}^{\alpha'}\psi_i^{\beta'} \left(\frac{\hat{\psi}_{i+1}^\dagger-\hat{\psi}_{i}^\dagger}{\epsilon}\right)\left(\frac{\hat{\psi}_{i+1}-\hat{\psi}_{i}}{\epsilon}\right)\psi_i^{\dagger \alpha}\psi_{i+1}^{\dagger \beta}|\Omega\rangle
 \end{eqnarray*}
as a series in $\epsilon$; the term $[Q,R]\otimes [\bar{Q},\bar{R}]$ is the only term independent of $\epsilon$.

Let us next illustrate how these continuous MPS can be used as a variational ansatz for strongly correlated continuous theories by applying them on the Lieb-Liniger model \cite{LiebLiniger}. The Lieb-Liniger Hamiltonian describes (non-relativistic) bosons in 1 spatial dimension interacting via a contact potential:
\[\mathcal{H}= \int_{-\infty}^{+\infty} dx \left[ \frac{d \hat{\psi}^\dagger(x)}{dx} \frac{d \hat{\psi}(x)}{dx} + c \hat{\psi}^\dagger(x) \hat{\psi}^\dagger(x) \hat{\psi}(x) \hat{\psi}(x)\right] \]
It is well known that, in the limit  $L\rightarrow\infty$, the energy density in this system can be expressed as $E/L=\rho^3 e(c/\rho)$ with $\rho$ the density and $e(c)$ the energy density of the system at $\rho=1$. This scaling can also readily be understood from the continuous MPS ansatz: for $L\to\infty$, $e^{LT}$ remains invariant under the scaling transformation $Q\to xQ$ and $R\to \sqrt{x} R$. Since density, kinetic, and interaction energy behave like $R\times R$, $[Q,R]\otimes [Q,R]$, and $R^2\otimes R^2$, respectively, we have that under this transformation $\rho\to x \rho$, $E_{kin}\to x^3 E_{kin}$ and $E_{int}\to x^2 E_{int}$. Thus,
$E_{kin}(\rho) + 2 c E_{int}(\rho) = \rho^3 \left(E_{kin}(\rho=1) + c/\rho E_{int}(\rho=1)\right)$ giving the above scaling. The energy density $e(c)$ can be determined in terms of the Bethe ansatz \cite{LiebLiniger}, whereas other quantities whereas other quantities like correlation functions have been calculated using Monte Carlo methods Ref. \cite{As,As2}.

\begin{figure}[t]
  \centering
\includegraphics[width=\linewidth]{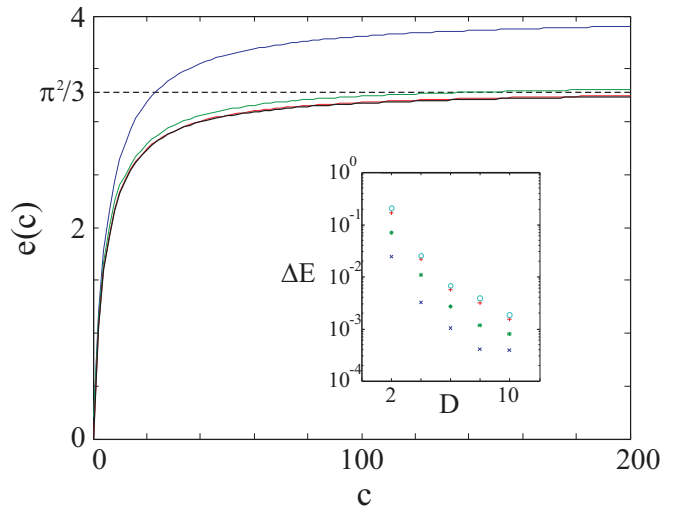}
  \caption{Energy density as a function of the interaction parameter $c$ for
different values of $D=2,4,6,8$ (from top to bottom). The result for
$D=8$ is indistinguishable from the one given by the Bethe Ansatz.
The insert shows the relative error $\Delta E=(e-e_{\it Bethe})/e_{\it Bethe})$ (where $E_{\it Bethe}$ is the energy given by the Bethe Ansatz solution), as
a function of $D$ for $c=0.2,2,20$ and $200$ (x,*,+, and o,, respectively).
We show the results for up to $D=10$; the saturation of the accuracy with $D=10$ is due to insufficient convergence of the results.}
\end{figure}

\begin{figure}[t]
  \centering
\includegraphics[width=\linewidth]{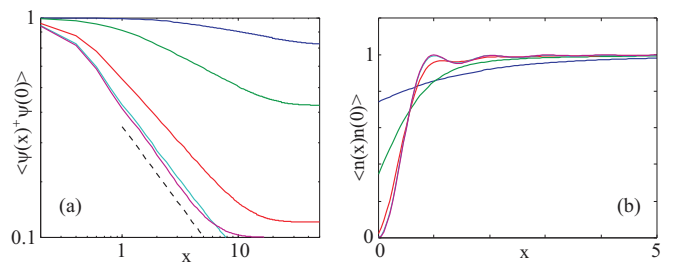}
  \caption{(a) Off-diagonal elements of the one-particle reduced density
operator as a function of the distance in a logarithmic scale for (from
top to bottom): $c=0.2,2,,20,200$ and $c=2000$. For reference we
have also drawn a straight line with slope=1/2, which is the slope
corresponding to the Tonk-Girardeau limit ($c\to\infty$). As it can be seen,
the slope of the curves approaches $1/2$ as $c$ increases. (b) Two-body
density-density correlation function for the same values of $c$ (in the left,
from bottom to top). For large $c$ one can observe the Friedel oscillations
corresponding to the Tonks-Girardeau limit.
All the results have been calculated with $D=14$.}
\end{figure}

We did a variational optimization of cMPS as a function of the scaling parameter $c$ (i.e., we chose $\rho=1$). We  carried out a simple gradient minimization of the energy density as a function of the matrices $A=i\tilde H$ and $R=O D$, where $A$ is antisymmetric, $O$ orthogonal and $D$ diagonal. In Fig. 1 we have plotted $e(c)$ for different values of the bond dimension $D$, as well as the one obtained by Bethe ansatz; The insert shows the relative error in the energy as a function of $D$, which seems to indicate an exponential dependence. As a comparison, for $c=2$ the Bethe ansatz gives $e=1.0504$, the Monte Carlo method of Ref. \cite{As,As2} gives $e=1.0518$, whereas we obtain $e=1.1241, 1.0618,1.0531,1.0515,1.0512$, and $1.0508$ for $D=2,4,\ldots,12$. In Fig. 2 we have determined the one-particle and density-density correlation functions. With  little numerical effort we obtain results which are comparable to those of exact Monte Carlo methods. By using more sophisticated techniques to perform the minimization we believe that much more precise results can be obtained, and thus cMPS can be viewed as an alternative to other existing methods \cite{Caux,Calabrese}. Importantly, the cMPS method does not rely on the fact that the model is integrable, and works equally well for non-intregrable models.

Let us next comment on how to do the calculations in the  case  the translational invariance is broken. This is obviously of central importance for the simulation of atomic gasses in a non-homogeneous potential such as occurring in optical lattices; the present ansatz allows to deal with the full Hamiltonian as opposed to  effective Hamiltonians such as the Bose-Hubbard model which typically ignore the potentially important effects from the higher Hubbard bands.  In that case, one should expand the functionals $Q(x),R(x)$ as a series, in such a way that a discrete amount of parameters characterize the state:
\[Q(x)=\sum_{k=0}^p f_p(x) Q_p\hspace{.5cm} R(x)=\sum_{k=0}^p f_p(x) R_p\]
Here the functions $f_p(x)$ can be chosen to correspond to harmonic Fourier functions  in the case of a periodic lattice or by localized functions in the case of e.g. a harmonic trap. For periodic lattices, this leads to a Bloch-like ansatz, and it is possible to define eigenfunctions of the site-dependent Lindblad operator in terms of a similar Fourier series. Similarly, it is possible to incorporate the MPS techniques for real-time evolution. In this case, the $Q(x,t)$ and $R(x,t)$ become both functions of space and time, and it is possible to write down coupled differential equations that describe the evolution.

Other obvious extensions include the simulation of systems with different types of fermions and/or bosons. This is relevant for the case of the Hubbard type models, where there are 2 types of fermions per site or in the case of mixtures. In this case, the  cMPS ansatz becomes
\[{\rm Tr}_{aux}\left[\mathcal{P}\exp\left(\int_0^L Q(x)\otimes\openone+\sum_\alpha R_\alpha(x)\otimes \hat{\psi}^{\dagger}_\alpha(x) \right)\right]|\Omega\rangle\]
where the $\psi_\alpha$ are field operators corresponding to different spins
(or species). Obviously, more local terms can be added in the exponential, such as
 \[ \sum_{\alpha\beta} S _{\alpha\beta}(x)\otimes \hat{\psi}_\alpha^\dagger(x)\hat{\psi}^{\dagger}_\beta(x) + S _{\alpha\beta}'(x)\otimes \hat{\psi}_\alpha(x)\hat{\psi}^{\dagger}_\beta(x).\]
Besides that, it is possible to extend this formalism to 2-dimensional continuum systems using the formalism of PEPS \cite{PEPS}. In that case, the auxiliary bond dimension has to be interpreted as representing an auxiliary field, and the judicious choice of tensors $Q$ and $R$ allows to develop a consistent formalism for describing 2+1 dimensional field theories \cite{inpreparation}.

In conclusion, we have introduced a new family of states, the cMPS, for quantum field models in 1 spatial dimension. They correspond to the continuum limit
of the MPS.  We have shown how one can efficiently
determine expectation values of different observables, so that they can be used to approximate ground state of such systems. There are many possible extensions of the present work. On the one hand, one can apply the same techniques as with MPS to describe mixed states or systems at finite temperature, as well as higher dimensions \cite{reviewMPS}. On the other hand, it would be interesting to explore new methods for finding the matrices $Q$ and $R$ variationally with high bond dimension, as well as to study non-translationally invariant systems. Beyond that, it would also be interesting to substitute those matrices by operators acting on an infinite-dimensional Hilbert space as in \cite{German} in order to capture critical phenomena and to study relativistic quantum field theories. Finally, the cMPS formalism allows to construct Hamiltonians whose exact ground states are known, which leads to new solvable field theories \cite{inpreparation}.

\acknowledgements
This work was supported by the EU Strep project QUEVADIS, the ERC grant QUERG,  the FWF SFB grants FoQuS and ViCoM, and the DFG-Forschergruppe 635.


\begin{thebibliography}{99}

\bibitem{Wilson}
K.G. Wilson
{\it Rev. Mod. Phys.}
{\bf 47}
773 (1975)


\bibitem{White} S.R. White
{\it Phys. Rev. Lett.}
{\bf 69}
2863 (1992);
U. Schollw{\"o}ck
{\it Rev. Mod. Phys.}
{\bf 77}
259 (2005)


\bibitem{MPS} M. Fannes, B. Nachtergaele, R.F. Werner, {\it Commun. Math. Phys.} {\bf 144}, 443 (1992); S. Ostlund and S. Rommer, {\it Phys. Rev. Lett.} {\bf 75}, 3537 (1995); F. Verstraete, D. Porras, and J. I. Cirac, {\it Phys. Rev. Lett.} {\bf 93}, 227205 (2004).


\bibitem{reviewMPS} F. Verstraete, J.I. Cirac, V. Murg, {\it Adv. Phys.} {\bf  57}, 143 (2008); J.I. Cirac, F. Verstraete, {\it J. Phys. A: Math. Theor.} {\bf  42},  504004  (2009)



\bibitem{VC06}
F. Verstraete and J.I.  Cirac
{\it Phys. Rev. B}
{\bf 73}
094423 (2006)

\bibitem{HastingsAREA}
M.B. Hastings
{\it J. Stat. Phys.}
P08024 (2007)

\bibitem{dynamicsMPS} G. Vidal
{\it Phys. Rev. Lett.}
{\bf 93}
040502 (2004); S.R. White and A.E.  Feiguin
{\it Phys. Rev. Lett.}
{\bf 93}
076401 (2004);
A.J. Daley, C. Kollath, U. Schollw{\"o}ck and G. Vidal
{\it J. Stat. Mech.: Theor. Exp.}
P04005 (2004);
F. Verstraete, J.J.  Garc{\'i}a-Ripoll, and J.I.  Cirac,
{\it Phys. Rev. Lett.}
{\bf 93}
207204
(2004)

\bibitem{PEPS} F. Verstraete  and J.I. Cirac
arXiv:cond-mat/0407066; see also G. Sierra and M.A. Martin-Delgado,  cond-mat/9811170.

\bibitem{LiebLiniger} E. H. Lieb and W. Liniger,
{\it Phys. Rev.} {\bf 130}, 1605 (1963)

\bibitem{inpreparation}
F. Verstraete and J. I. Cirac, in preparation.

\bibitem{As}
G.E. Astraharchik and S. Giorgini, {\it Phys. Rev. A} {\bf 68}, 031602 (2003)

\bibitem{As2}
G.E. Astraharchik and S. Giorgini, {\it J. Phys. A} {\bf 39}, 1 (2006)

\bibitem{Caux}
J.S. Caux, arXiv:0908.1660.

\bibitem{Calabrese}
J. S. Caux and P. Calabrese, {\it Phys. Rev. A} {\bf 74}, 031605 (2006).

\bibitem{German}
J. I. Cirac and G. Sierra, arXiv:0911.3029

\end{thebibliography}
\end{document}